\documentclass[12pt,a4paper]{article}
\usepackage{graphicx}
\usepackage{times}
\usepackage{enumitem}
\title{\bf An F-type ultra-fast rotator KIC\,6791060: Starspot modulation and Flares}
\author{Subhajeet Karmakar\thanks{subhajeet09@gmail.com, subhajeet@aries.res.in} ~and Jeewan C. Pandey\\
\vspace{1cm}\\
\normalsize Aryabhatta Research Institute of observational sciencES (ARIES), Nainital-263002, India}
\date{\mbox{}}
\begin{document}
\maketitle
\pagestyle{empty}
%
%
\def\bull{\vrule height .9ex width .8ex depth -.1ex}
\makeatletter
\def\ps@plain{\let\@mkboth\gobbletwo
\def\@oddhead{}\def\@oddfoot{\hfil\scriptsize\bull\quad
"First Belgo-Indian Network for Astronomy \& astrophysics (BINA) workshop'', held in Nainital (India), 15-18 November 2016 \quad\bull}%
\def\@evenhead{}\let\@evenfoot\@oddfoot}
\makeatother
%
%
\def\beginrefer{\section*{References}%
\begin{quotation}\mbox{}\par}
\def\refer#1\par{{\setlength{\parindent}{-\leftmargin}\indent#1\par}}
\def\endrefer{\end{quotation}}
%
%
\def\kepler{{\it Kepler}}

{\noindent\small{\bf Abstract:} 
Making use of the data obtained from \kepler\ satellite, we have analysed an F-type ultra-fast rotator KIC\,6791060. We derive a rotational period of 0.34365$\pm$0.00004 d. Multiple periodicity with a period separation of $\sim$0.00016 d was detected, which appears to be a result of the relative velocity between the multiple spot-groups in different stellar latitudes due to surface differential rotation. During the observations, 38 optical flares are detected. Energies of these flares have been found to be in the range of $10^{31-33}$ erg. Preliminary results of surface temperature modeling indicate the existence of two active longitudes on the stellar surface.
}
%
%
\section{Introduction}
Stars with a spectral type from late-F to early-K have a convective envelope
above a radiative interior with an interface where strong shear leads to
amplification of magnetic fields. The observational evidence of magnetic 
activities are surface inhomogeneities, short and long-term variations in 
spot cycles, and flares.
These magnetic activities are even more interesting when the
rotational velocities are very high ($>$40 km s$^{-1}$). We have chosen a poorly known F-type, main sequence, ultra-fast rotator (UFR) KIC\,6791060 with a period of 0.344 d (Balona 2015), $V$-band magnitude of 10.57 mag (H{\o}g et al. 2000), and $B-V$ color of 0.44 mag (H{\o}g et al. 2000) with the aims to study the short period cycles, surface inhomogeneities, surface differential rotation (SDR) and flares.

\section{\kepler\ observations and data processing}
\begin{figure}[t]
\centering
\includegraphics[width=14cm]{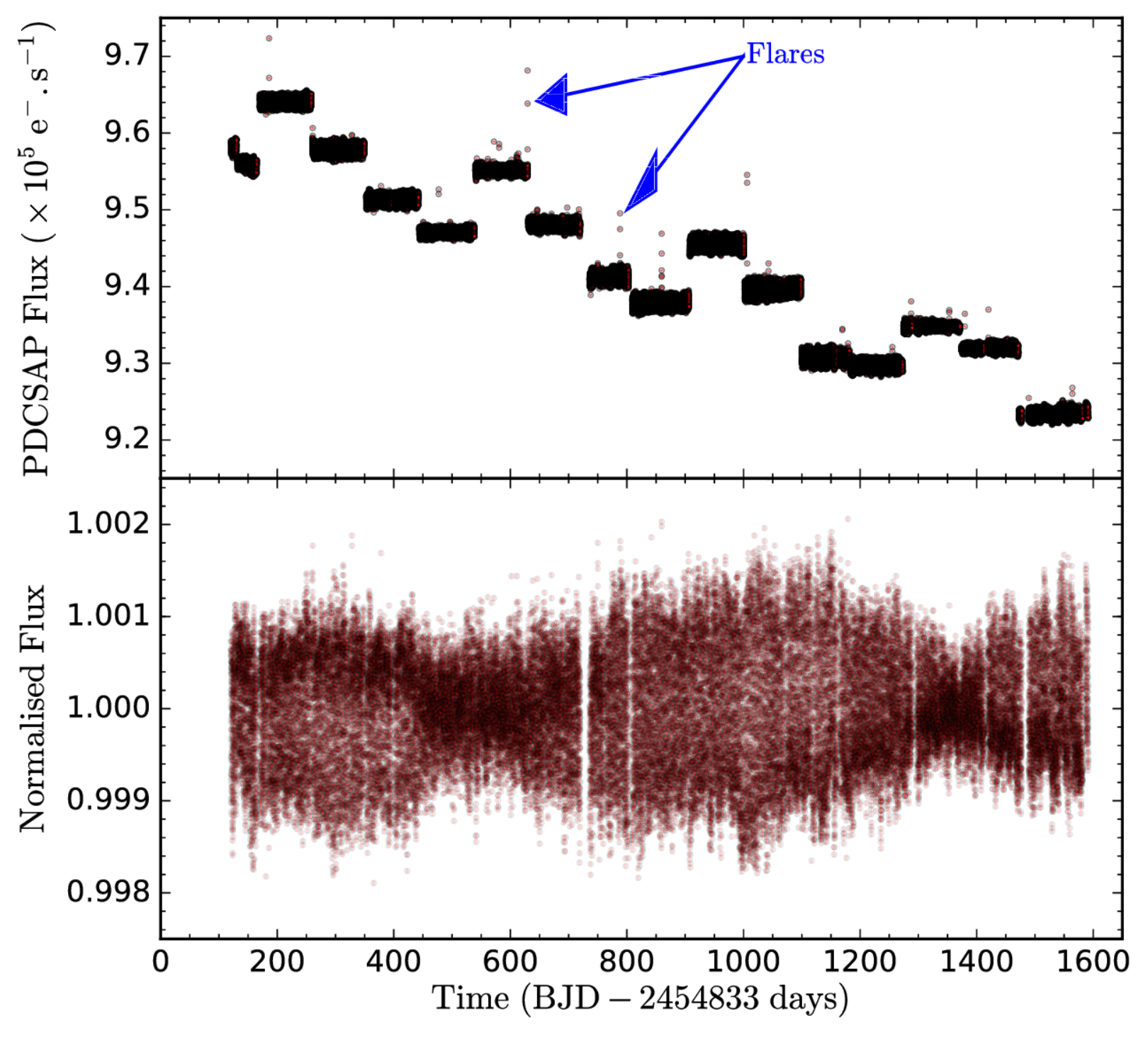}
\caption{ Top: The raw PDCSAP light curve for 18 quarters of LC mode data for KIC\,6791060.
Bottom: The flare/outlier removed light curve after additional linear corrections. \label{fig_1}}
\end{figure}

The F-type star KIC\,6791060 was observed for four years from 13 December 2009 to 13 December 2013 by the \kepler\ satellite. The Kepler data consist of practically continuous photometry for that period. The data were subdivided into eighteen quarters with 30-min exposures (long-cadence; LC mode). In one quarter only it was observed with 1-min exposures (short-cadence; SC mode). For our analysis we used the \textit{pre-search data conditioning simple aperture photometry} (PDCSAP) data in which instrumental effects are removed. The raw PDCSAP light curve obtained in LC mode is shown in upper panel of Fig.~\ref{fig_1}. For further analysis, we have subtracted a linear fit from the light curve of each individual quarter to remove any long-term calibration errors and normalised it to its average flux.

\section{Light curves and period analysis}
\begin{figure}[t]
\begin{minipage}{8cm}
\centering
\includegraphics[width=8.5cm]{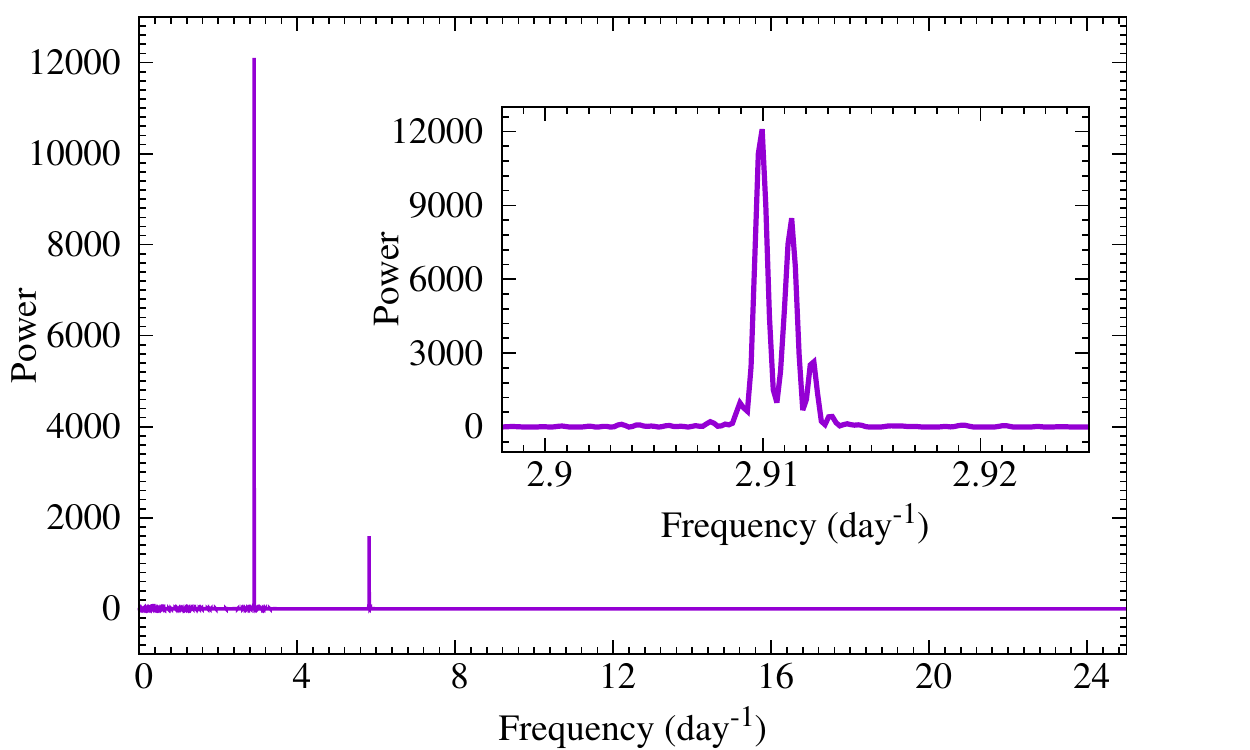}
\caption{Scargle-Press periodograms obtained from the Kepler light curve
of KIC\,67091060. The insets shows an expanded view on the strongest peaks. \label{fig_2}}
\end{minipage}
\hfill
\begin{minipage}{8.5cm}
\centering
\includegraphics[width=8.5cm]{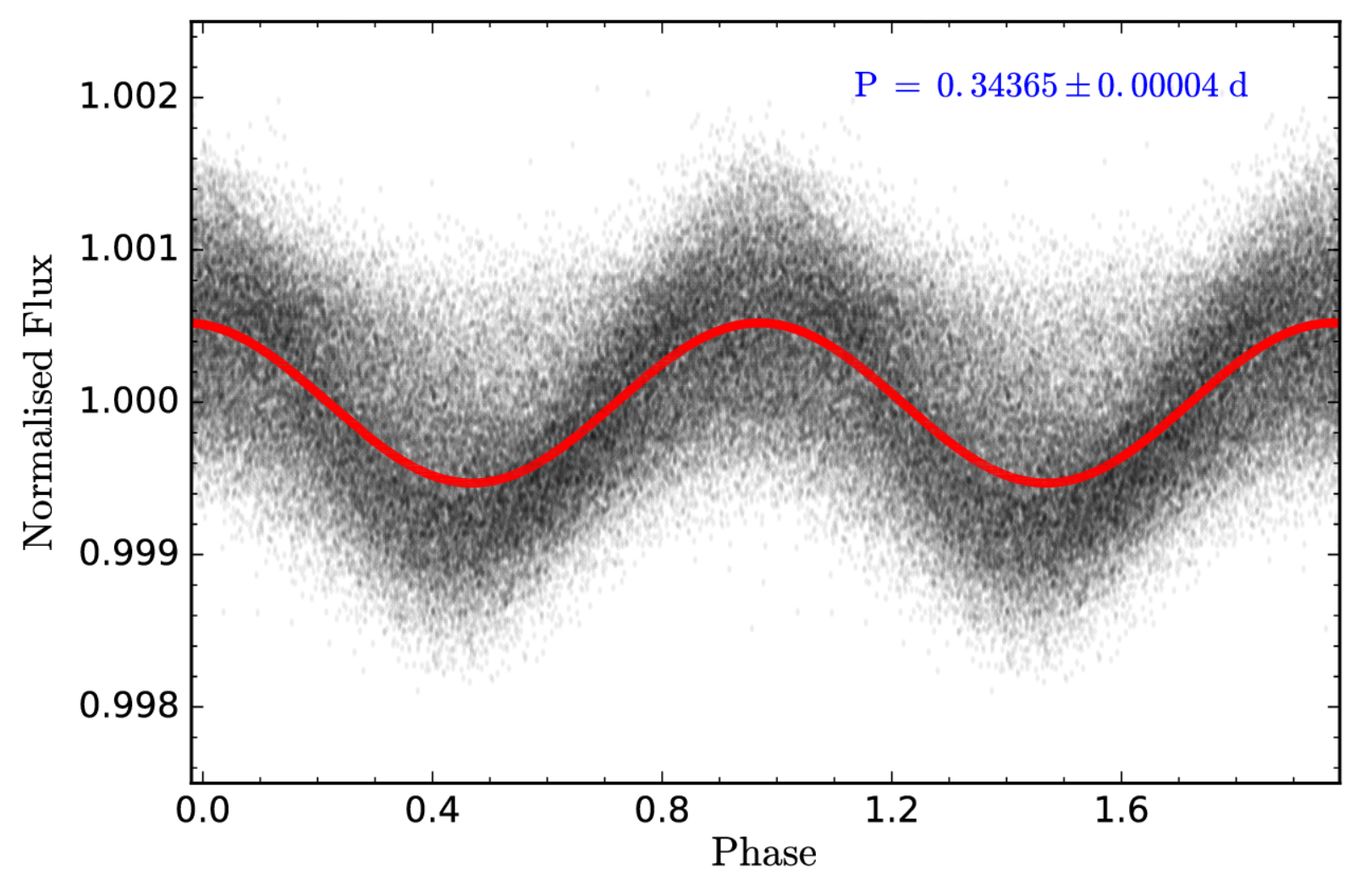}
\caption{Folded light curve of KIC\,67091060 with the period that corresponds to the strongest peak in Scargle-Press periodogram (see value in the upper right corner). The best fitting sinusoid is shown by the continuous line. The density of the data shows the spread in amplitude during these four years of observation. \label{fig_3}}
\end{minipage}
\end{figure}

The flare/outlier-removed, flat, normalised light curve is shown in the bottom panel of Fig.~\ref{fig_1}. The light curve displays both short- and long-term flux variations with a high amplitude. We searched for periodicities in the data using the Scargle-Press period search method (Scargle 1982; Horne \& Baliunas 1986; Press \& Rybicki 1989). The Scargle power spectrum is shown in Fig.~\ref{fig_2}. The highest peak in the power spectrum corresponds to a period of 0.34365$\pm$0.00004 d, which could be the rotation period of the star. Our derived period agrees with previous determinations for KIC\,6791060 (Balona 2015, McQuillan et al. 2014). 
 In our study, another peak was also seen in the power spectrum which corresponds to a period of 0.34349$\pm$0.00004 d, and is separated by 0.00016 d from the rotational period. 
This multiple periodicity throughout the observations results in a beat pattern (minimal amplitude near $\sim$550 and $\sim$1350 days in time = BJD - 2\,454\,833 days; see bottom panel of Fig.~\ref{fig_1}). According to our preliminary analysis, this multiple periodicity could be a result of multiple spot groups located in different latitude while the difference in period could be due to the SDR of the star. Fig.~\ref{fig_3} shows the phase-folded light curve of KIC\,6791060, folded with the dominant frequency in the power spectrum, where the continuous line is the best fitting sinusoid. This indicates that the derived periodicity is real and shows a sinusoidal variation. 

%
%

\section{Flare analysis}
\begin{figure}
\centering
\includegraphics[height=13cm,angle=-90]{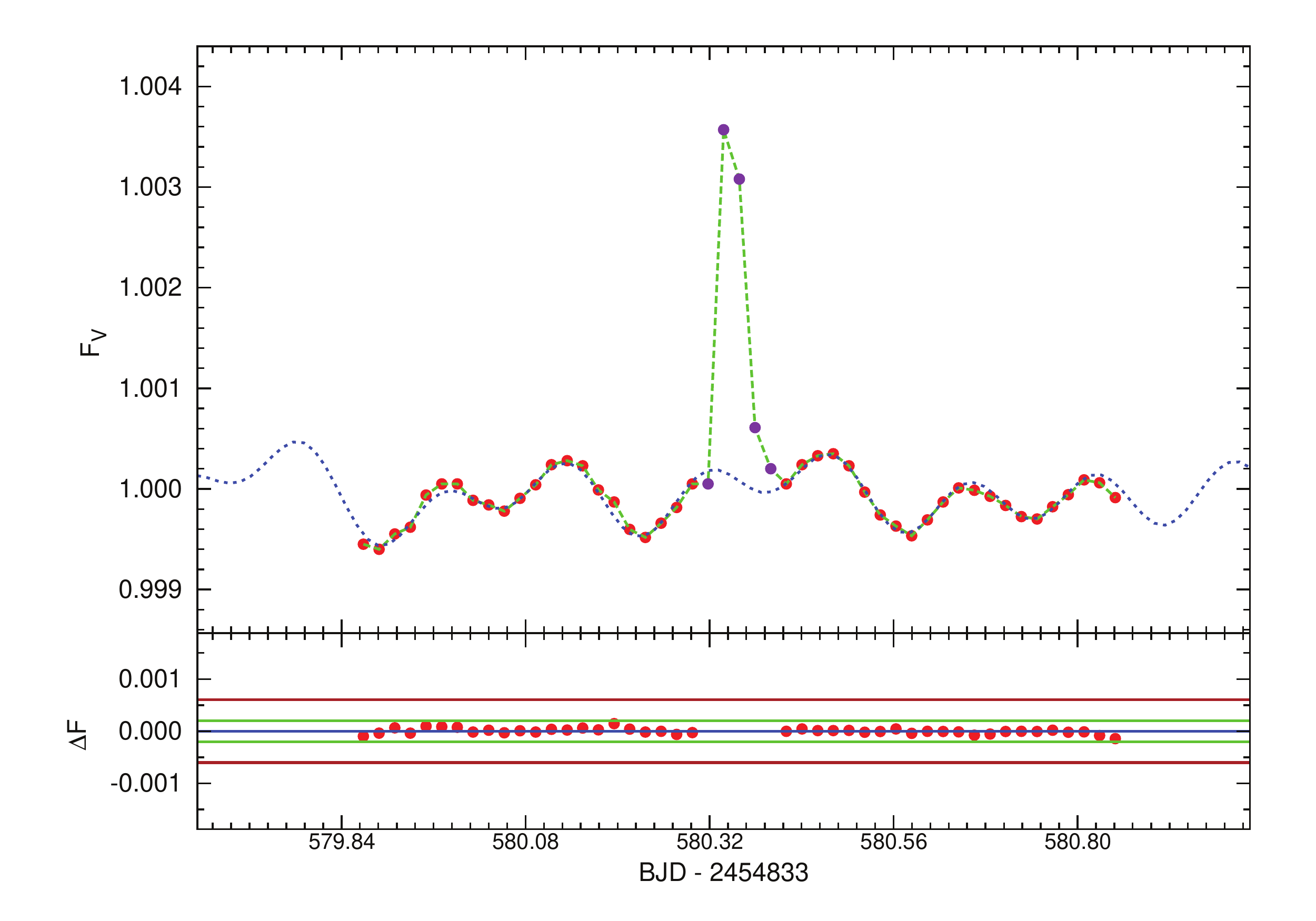}
\caption{Light curve of a sample flare on KIC\,6791060. The dotted line in top panel shows the best-fit sinusoid along with the light curve. The bottom panel shows the detrended light curve (excluding flare region) along with the 1$\sigma$ and 3$\sigma$ level. \label{fig_4}}
\end{figure}

Flares are apparently visible as positive flux excursions throughout the data (see top panel of Fig.~\ref{fig_1}).
In our flare identification method, we took great care so that we do not miss any small flares, and that we do not misdetect an outlier as a flare either. In  order to do this job, we have used the following approach:
we have computed the moving standard deviation ($\sigma$) of the detrended light curve with a data-length equal to twice the rotational period of the star. The data points in any time segments deviating from the detrended light curve gives a larger $\sigma$-value and they were flagged as flare candidates. The light curves are detrended by fitting the sinusoids with periods as obtained from period analysis.
All the flagged flare candidates were verified manually and identified as flare only if it contains 3 consecutive points above the 3$\sigma$ level.
Fig.~\ref{fig_4} shows the light curve of a sample flare on KIC\,6791060. The dotted line in top panel shows the best-fit sinusoid, whereas the bottom panel shows the detrended, flare-excluded light curve along with the 1$\sigma$ and 3$\sigma$ level. A total of 38 flares were detected in KIC\,6791060 with a flare rate of 1 flare per 40 days. The flare frequency is much lower than the one of G--M dwarfs. This indicates that F-dwarfs are less active than G--M dwarfs and other active UFRs like LO Peg (Karmakar et al. 2016) and AB Dor (Vilhu et al. 1993). 

The flare energy is computed using the area under the flare light curve
i.e. the integrated excess flux ($F_{e}(t)$) released during the flare (from flare start-time $t_s$ to flare end-time $t_e$) as
\begin{equation} 
E_{flare} = 4\pi d^{2} \int_{t_s}^{t_e} \! F_{e}(t) \, dt
\label{eqn:energy}
\end{equation}

With a distance (d) of 79 pc (Ammons et al. 2006) for KIC\,6791060 the
derived flare energies were found in the order of 10$^{31-33}$ erg. These flare energies are similar to those of solar flares, but smaller than the ones of other active UFRs like LO Peg (Karmakar et al. 2016).

\section{Summary}
\begin{itemize}
  \item Using the Scargle-Press periodogram, we found the rotational period of the F-type UFR KIC\,6791060 to be 0.34365$\pm$0.00004 d.
  \item KIC\,6791060 seems to have multiple spot-groups at different latitude over the period of observation.
  \item A total 38 flares are identified on KIC\,6791060, with a flare rate of one flare per 40 days. 
  \item The flare energies are found to be of the order of 10$^{31-33}$ erg.
\end{itemize}

\vspace{-0.5cm}
\section*{Acknowledgements}
This paper includes data collected by the \textit{Kepler} mission.
Funding for the \textit{Kepler} mission is provided by the NASA
Science Mission directorate.
%
%
%


\footnotesize
\beginrefer
\refer Balona L. A. 2015, MNRAS, 447, 2714

\refer H{\o}g E., Fabricius C., Makarov V. V. et al. 2000, A\&A, 355, L27

\refer Horne J. H., Baliunas S. L. 1986, ApJ, 302, 757

\refer Karmakar S., Pandey J. C., Savanov I. S. et al. 2016, MNRAS, 459, 3112

\refer McQuillan A., Mazeh T., Aigrain S. 2014, ApJS, 211, 24

\refer Press W. H., Rybicki G. B. 1989, ApJ, 338, 277

\refer Scargle J. D. 1982, ApJ, 263, 835

\refer Vilhu O., Tsuru T., Collier Cameron A. et al. 1993, A\&A, 278, 467

\endrefer

\end{document}